\documentstyle[12pt,epsf]{article}
\newcommand{\beq}{\begin{equation}}
\newcommand{\eeq}{\end{equation}}
\newcommand{\beqa}{\begin{eqnarray}}
\newcommand{\eeqa}{\end{eqnarray}}
\newcommand{\ba}{\begin{array}}
\newcommand{\ea}{\end{array}}
\newcommand{\CR}{\nonumber \\}
\newcommand{\pa}{\partial}
\newcommand{\A}{\alpha}
\newcommand{\B}{\beta}
\newcommand{\D}{\delta}          

\newcommand{\E}{\epsilon}

\newcommand{\p}{\Phi}

\newcommand{\lm}{\lambda}

\newcommand{\la}{{\langle}}
\newcommand{\ra}{{\rangle}}
\newcommand{\half}{{1\over 2}}
\newcommand{\Tr}{{\rm Tr}}

\newcommand{\cO}{{\cal O}}
\newcommand{\cL}{{\cal L}}
\newcommand{\hF}{{\hat F}}
\newcommand{\hA}{{\hat A}}
\newcommand{\hD}{{\hat D}}
\newcommand{\T}{\theta}

\begin{document}

\makeatletter
\def\setcaption#1{\def\@captype{#1}}
\makeatother

\begin{titlepage}
\null
\begin{flushright} 
hep-th/0001111  \\
UT-871  \\
January, 2000
\end{flushright}
\vspace{0.5cm} 
\begin{center}
{\Large \bf
On the Equivalence between Noncommutative 
and Ordinary Gauge Theories
\par}
\lineskip .75em
\vskip2.5cm
\normalsize
{\large Seiji Terashima}\footnote{
E-mail:\ \ seiji@hep-th.phys.s.u-tokyo.ac.jp} 
\vskip 1.5em
{\large \it  Department of Physics, Faculty of Science, University of Tokyo\\
Tokyo 113-0033, Japan}
\vskip3cm
{\bf Abstract}
\end{center} \par
Recently Seiberg and Witten have proposed that 
noncommutative gauge theories realized as effective theories 
on D-brane are equivalent to
some ordinary gauge theories.
This proposal has been proved, however, 
only for the Dirac-Born-Infeld action
in the approximation of neglecting
all derivative terms.
In this paper we 
explicitly construct general forms of 
the $2 n$-derivative terms
which satisfy this equivalence under their assumption
in the approximation of 
neglecting $(2 n+2)$-derivative terms.
We also prove
that the D-brane action computed in the superstring theory
is consistent with the equivalence
neglecting the fourth and higher order derivative terms.

\end{titlepage}

\baselineskip=0.7cm

\section{Introduction}

\renewcommand{\theequation}{1.\arabic{equation}}\setcounter{equation}{0}

Recently it has been realized that
the noncommutative geometry has played a profound role 
in a specific compactification of the 
Matrix theory \cite{CoDoSc} and also in superstring theory 
via D-branes with constant $B$ fields
\cite{DoHu}-\cite{Sc}.
From deeper investigation of the noncommutative gauge theory via
D-branes,
Seiberg and Witten have proposed that
the noncommutative gauge theories realized as effective theories 
on D-branes are equivalent to
some ordinary gauge theories \cite{SeWi}.
In a single D-brane case, it has been known that 
the effective action on the brane is Dirac-Born-Infeld action 
if all derivative terms are neglected \cite{FrTs}-\cite{AbCaNaYo}.
Thus the Dirac-Born-Infeld action should be consistent 
with the equivalence in this approximation.
Indeed this has been shown in \cite{SeWi}.

It is a very natural question whether
the equivalence indeed holds beyond
the approximation of neglecting all derivative terms, or not.
If it holds without the approximation,
the forms of derivative corrections have to be highly restricted.
Moreover when this equivalence is strong enough,
we can determine the effective action completely from
it with the help of other requirements and
study the dynamics of the D-branes using the action.

In this paper, we show that in the approximation of 
neglecting the fourth and higher order derivative terms
the D-brane action computed in the superstring theory
is consistent with the equivalence.
Although, this is 
the Dirac-Born-Infeld action without two-derivative corrections,
to show the equivalence we should take into account 
the orderings of the noncommutative field strength 
in the Dirac-Born-Infeld action.
By taking appropriate ordering it becomes consistent
and this is regarded as a non-trivial test of the equivalence.

With the mapping of the ordinary
gauge field to noncommutative gauge field given in \cite{SeWi},
we also explicitly construct 
general forms of the two-derivative corrections
which satisfy the equivalence relation 
in the approximation of 
neglecting the four-derivative terms.
Furthermore, we can construct 
the $2n$-derivative corrections
which are consistent with the equivalence
in the approximation of neglecting the $(2n+2)$-derivative terms.
It should be emphasized that the results obtained in this paper are
valid for arbitrary order of the field strength.\footnote{
In this paper, we regard $F_{ij} \sim {\cO} ( \pa^{0})$ 
and $A_i \sim {\cO} ( \pa^{-1})$  }

On the other hand, in \cite{Ok} it has been shown that 
for the bosonic string case
the known two-derivative correction \cite{AnTs,AnTs2}
is not consistent with the equivalence.\footnote{
In \cite{Ok} the consistent two-derivative corrections 
up to the quartic order of the field strength have been considered.
As we will see later, the result obtained in this paper
reproduce these corrections .}
This problem can be resolved by considering
$B$-dependent field redefinition of the $U(1)$ gauge field 
\cite{OkTe}.
Therefore in order to constrain the effective action
for the bosonic string case,
we should include the 
$B$-dependent field redefinition.
However two-derivative corrections allowed
by the equivalence
without the $B$-dependent field redefinition
may also be allowed by the equivalence with it \cite{OkTe}.
Furthermore,
the higher derivative corrections 
may capture some general structures of the 
effective action of the D-brane.
Therefore the derivative corrections obtained in this paper
are probably important.

This paper is organized as follows.
In section 2, we briefly 
review the equivalence between noncommutative 
and ordinary gauge theories shown in \cite{SeWi}.
In section 3 
it is shown that the certain noncommutative version
of the DBI actions without two-derivative corrections
are consistent with the equivalence 
in the approximation of neglecting the four-derivative terms.
We also construct the consistent two-derivative corrections 
in this approximation.
In section 4 we argue that 
the two-derivative corrections obtained in section 3 
exhaust the consistent two-derivative corrections and also
generalize these to the $2n$-derivative corrections.
Finally section 5 is devoted to conclusion.

\section{Noncommutative Gauge Theory}

\renewcommand{\theequation}{2.\arabic{equation}}\setcounter{equation}{0}

In this section we briefly 
review the equivalence between noncommutative 
and ordinary gauge theories shown in \cite{SeWi}.
We consider open strings in flat space, with metric $g_{ij}$,
in the presence of a constant $B_{ij}$ and with a Dp-brane.
Here we assume that $B_{ij}$ has rank $p+1$ and $B_{ij} \neq 0$
only for $i,j=1, \ldots, p+1$.
The world-sheet action is
\beq
S=\frac{1}{4 \pi \A'} \int_{\Sigma} g_{ij} \pa_a x^i \pa^a x^j
-\frac{i}{2} \int_{\pa \Sigma} B_{ij} x^i \pa_{\tau} x^j
-i \int_{\pa \Sigma} A_i(x) \pa_{\tau} x^i,
\eeq
where $\Sigma$ is the string world-sheet, $\pa_{\tau}$ is
the tangential derivative 
along the world-sheet boundary $\pa \Sigma$ and
$A_i$ is a background gauge field.
In the case that $\Sigma$ is the upper half plane parameterized by 
$-\infty \leq \tau \leq \infty$ and $0 \leq \sigma \leq \infty$, 
the propagator evaluated at boundary points is \cite{FrTs}-\cite{AbCaNaYo}
\beq
\la x^i (\tau) x^j (\tau')  \ra = -\A' (G^{-1})^{ij} \log (\tau-\tau')^2+
\frac{i}{2} {\T}^{ij} \E (\tau-\tau'),
\eeq
where $G$ and $\T$ are 
the symmetric and antisymmetric tensors defined by
\beq
(G^{-1})^{ij} +\frac{1}{2 \pi \A'} \T^{ij} =
\left( \frac{1}{g+2 \pi \A' B} \right)^{ij}.
\eeq

From considerations of the string S-matrix,
the $B$ dependence of the effective action for fixed $G$ 
can be obtained by replacing ordinary multiplication 
in the effective action for $B=0$ by the $*$ product 
defined by the formula
\beq
\left.
f(x)*g(x)=e^{\frac{i}{2} \T^{ij}
\frac{\pa}{\pa \xi^i} \frac{\pa}{\pa \zeta^j} } f(x+\xi) g(x+\zeta )
\right|_{\xi=\zeta=0}.
\eeq
It is likely that the gauge transformation also 
becomes noncommutative.
In fact, using the point splitting regularization,
$S$ is invariant under noncommutative gauge transformation
\beq
\hat{\D} \hA_i=\hD_i \lm,
\eeq
where covariant derivative $\hD_i$ is defined as
\beq
\hD_i E(x)= \pa_i E(x)+ i  \left( E(x) * \hA_i -\hA_i * E(x) \right).
\eeq

Conversely, using Pauli-Villars regularization, 
$S$ is invariant under ordinary gauge transformation
\beq
\D A_i=\pa_i \lm.
\eeq
Therefore, the effective Lagrangian obtained in this way becomes
ordinary gauge theory.
Thus this theory and the corresponding noncommutative gauge theory
are equivalent under the field redefinition $\hA=\hA(A)$ 
since the coupling constants in the world-sheet theory 
are the spacetime fields.
Because the two different gauge invariance should satisfy
$\hA(A)+\hat{\D}_{\hat{\lm}} \hA(A)=\hA(A+\D_{\lm} A)$,
the mapping of $A$ to $\hA$ for $U(1)$ case 
is obtained as a differential equation
for $\T$,
\beqa
\D \hA_i(\T) = \D \T^{kl} \frac{\pa}{\pa \T^{kl} }\hA_i(\T)
&=& -\frac{1}{4}\D \T^{kl} 
\left[ \hA_k *(\pa_l \hA_i+\hF_{li})+(\pa_l \hA_i+\hF_{li})*\hA_k  ] 
\right. \CR 
\D \hF_{ij}(\T) = \D \T^{kl} \frac{\pa}{\pa \T^{kl} }\hF_{ij}(\T)
&=& \frac{1}{4}\D \T^{kl} [ 2 \hF_{ik}* \hF_{jl}+2 \hF_{jl}* \hF_{ik} 
\CR 
&& \hspace{.5cm} 
 - \hA_k *\left( \hD_l \hF_{ij} + \pa_l \hF_{ij} \right)
-\left( \hD_l \hF_{ij} + \pa_l \hF_{ij} \right)* \hA_k ],
\label{map}
\eeqa
where
\beq
{\hF}_{ij}=\pa_i \hA_j -\pa_j \hA_i-i \hA_i * \hA_j+i \hA_j * \hA_i.
\eeq
In \cite{Oku} this map has been derived in a path integral form 
from D-brane world-volume perspective \cite{Is,CoSc}.\footnote{
Although the differential equation has ambiguities \cite{AsKi}, 
these ambiguities have no physical meaning
because they correspond to the field redefinition.}

In the approximation of 
neglecting the derivative terms, 
the effective Lagrangian is the 
Dirac-Born-Infeld Lagrangian
\beq
\cL_{DBI}=\frac{1}{g_s (2 \pi)^p (\A')^{\frac{p+1}{2}}}
\sqrt{\det(g+2 \pi \A' (B+F))},
\eeq
where $F_{ij}=\pa_i A_j -\pa_j A_i$.
Here $g_s$ is the closed string coupling and 
the normalization of the Lagrangian is same 
as the one taken in \cite{SeWi}.
Therefore the equivalent noncommutative gauge theory
in the approximation has the following Lagrangian
\beq
\hat{\cL}_{DBI}=\frac{1}{G_s (2 \pi)^p (\A')^{\frac{p+1}{2}}}
\sqrt{\det(G+2 \pi \A' \hF)}.
\label{Ln}
\eeq
Note that all the multiplication entering the 
r.h.s of (\ref{Ln}) can be regarded as 
the ordinary multiplication except those in the definition of $\hF$
because of the approximation.
From the requirement $\cL_{DBI}=\hat{\cL}_{DBI}$ for $F=0$,
the overall normalization $G_s$ should be fixed 
as $G_s=g_s \sqrt{\det(G)/ \det(g+2 \pi \A' B)}$.

Furthermore, in \cite{SeWi} it has been proposed that
the effective action can be written for arbitrary values of $\T$.
More precisely 
for given physical parameters $g_s, g_{ij}$ and $B_{ij}$ and
an auxiliary parameter $\T$,
we define $G_s, G_{ij}$ and a two form $\p_{ij}$ as
\beqa
\left( \frac{1}{G+2 \pi \A' \p} \right)^{ij} &=&
-\frac{1}{2 \pi \A'} \T^{ij} 
+\left( \frac{1}{g+2 \pi \A' B} \right)^{ij} \CR
G_s &=& g_s \left(
\det \left(-\frac{1}{2 \pi \A'} \T +\frac{1}{g+2 \pi \A' B} \right)  
\det(g+2 \pi \A' B) \right)^{-\half}.
\eeqa
Then the effective action $\hat{S}(G_s,G,\p,\T ; \, \hF)$, in which
the multiplication is the $\T$-dependent $*$ product,
is actually $\T$-independent, i.e. 
$\hat{S}(G_s,G,\p,\T ; \, \hF)=S(g_s,g,B,\T=0; \, F)$.
The effective action including $\p$ may be 
obtained using a regularization which
interpolates between Pauli-Villars and point splitting as in \cite{AnDo}.
In this paper, we simply assume
this proposal.

In the approximation of neglecting the derivative of $F$,
the equation 
\beq
\D \cL_{\p}= 
\left. \D \T^{kl} \frac{\pa \cL_{\p} }{\pa \T^{kl}} 
\right|_{g_s,g,B,A_i \, fixed}
= {\rm total} \,\, {\rm derivative}
+{\cO} (\pa^2),
\label{phiLd}
\eeq
should hold.\footnote{
Hereafter $\D$ always denotes $\D \T^{kl} \frac{\pa }{\pa \T^{kl}}$.}
Here $\cL_{\p}$ is the Lagrangian defined as
\beq
\cL_{\p}=\frac{1}{G_s (2 \pi)^p (\A')^{\frac{p+1}{2}}}
\sqrt{\det(G+2 \pi \A' (\hF+\p))},
\label{phiL}
\eeq
where the multiplication is the $*$ product except in the definition of $\hF$.
Below for simplicity we set $2 \pi \A'=1$.
The variation of $G_s, G$ and $\p$ are 
\beqa
\D G_s &=& \frac{1}{2} G_s \Tr ( \p \D \T ), \CR
\D G &=& G \D \T \p + \p \D \T G, \CR
\D \p &=& \p \D \T \p + G \D \T G,
\eeqa
and the variation of $\hF$ is
\beqa
\D \hF_{ij} &=& -( \hF \D \T \hF)_{ij} -\hA_k \D \T^{kl} 
\frac{1}{2} (\pa_l+\hD_l) \hF_{ij} +\cO (\pa^4) \CR
&=& -( \hF \D \T \hF)_{ij} -\hA_k \D \T^{kl} 
(\pa_l -\half \T^{mn} \pa_n \hA_l \pa_m ) \hF_{ij} +\cO (\pa^4).
\eeqa

Following \cite{SeWi}, we get
\beqa
\!\!\!\!\!\!\!\!\! &&  \!\!\!\!\! 
\D \left( \frac{1}{G_s} \det (G+\hF+\p)^{\half} \right) \CR
\!\!\!\!\!\!\!\!\! &&\!\!  =-\half \frac{1}{G_s} \det (G+\hF+\p)^{\half} 
\left( \Tr (\hF \D \T ) +\!\! \left( \frac{1}{G+\hF+\p} \right)_{ji} 
\!\! \hA_k \D \T^{kl} \half (\pa_l +\hD_l) \hF_{ij} \right) ,
\eeqa
where the multiplication is the ordinary one 
except in $\hF$ and $\hD_l$.
Now using
\beq
\half (\pa_l+\hD_l) \hA_k-
\half (\pa_k+\hD_k) \hA_l = 
\hD_l \hA_k -\pa_k \hA_l=\hF_{lk},
\eeq
we see that
\beqa
&& \!\!\!\!\! \D \T^{kl} (\pa_l+\hD_l) 
\left( \hA_k \det (G+\hF+\p)^{\half} \right) \CR
&& \hspace{0cm} =\D \T^{kl} \det (G+\hF+\p)^{\half} 
\left( \hF_{lk} +\!\!\half \!\!\left( \frac{1}{G+\hF+\p} \right)_{ji} 
\!\! \hA_k (\pa_l+\hD_l) \hF_{ij} \right)\!\! +\!\! \cO (\pa^4),
\eeqa
is a total derivative.
Thus we obtain the desired result
\beq
\D \left( \frac{1}{G_s} \det (G+\hF+\p)^{\half} \right)=
{\rm total} \,\, {\rm derivative} +\cO (\pa^4).
\label{comp}
\eeq
Note that the computation above shows
that the Dirac-Born-Infeld Lagrangian (\ref{phiL})
is $\T$-independent even in the approximation 
of neglecting $\cO (\pa^4)$ terms.

\section{Two-derivative terms}

\renewcommand{\theequation}{3.\arabic{equation}}\setcounter{equation}{0}

In this section, 
we will see that certain noncommutative Dirac-Born-Infeld actions
are consistent with the equivalence in the approximation
of neglecting four-derivative terms.
We also give certain two-derivative corrections
consistent with the
equivalence in the approximation of neglecting $\cO( \pa^4)$ terms.

Because the multiplication in 
the Dirac-Born-Infeld Lagrangian $\cL_{\p}$
should be replaced by the $*$ product
in the approximation,
we first consider the ordering of the $\hF_{ij}$ in 
the Dirac-Born-Infeld Lagrangian in which the multiplication 
is the $*$ product.
This Lagrangian is relevant in the approximation $\cO( \pa^4)$ 
and denoted as $\hat{\cL}_{\p}$.

It seems that there are two natural ways of ordering.
The first one is symmetrization of 
the $(\hF+\p)_{ij}$ in $\hat{\cL}_{\p}$,
as in the non-Abelian Dirac-Born-Infeld Lagrangian 
considered in \cite{Ts}.
The another one is as follows.
First we expand the square root of 
determinant using $U_{2n} \equiv \Tr (G^{-1}(\hF+\p))^{2n}$.
Next keeping the order of $(\hF+\p)_{ij}$ in $U_{2n}$ 
indicated by the symbol $\Tr$,
we symmetrize the polynomials of $U_{2n}$'s.
Then replacing all the multiplication by $*$ product,
we obtain the $\hat{\cL}_{\p}$ from the $\cL_{\p}$.

To take the either one of these, 
we will show 
\beq
\hat{\cL}_{\p}=\cL_{\p} + \cO( \pa^4).
\label{oo}
\eeq
To show this, we first remember 
the fact $f * g + g * f=2 fg +\cO (\pa^4)$.
Thus taking the first way of ordering, 
we easily see that (\ref{oo}) is satisfied
since $(\hF+\p)_{ij}$ is symmetrized.
If we take the second way, using
\beqa
\!\!\!\!\!\! && 
\Tr \left[ (G^{-1} (\hF+\p) ) *  \cdots *(G^{-1} (\hF+\p) ) \right]
\CR
\!\!\!\!\!\! &=& {\rm Tr} \left[ (G^{-1} (\hat{F}+\Phi))^{2n} \right] \CR
&&+ i \sum_{m=0}^{2 n-2} (m+1) \theta^{kl} 
{\rm Tr} \left[ (G^{-1} \partial_k \hat{F}) (G^{-1} (\hat{F}+\Phi) )^{2 n-m-2}
(G^{-1} \partial_l \hat{F} )  (G^{-1} (\hat{F}+\Phi) )^{m} \right] 
+{\cal O}( \partial^4) \CR
\!\!\!\!\!\! &=& \Tr \left[ (G^{-1} (\hF+\p))^{2n} \right] 
+\cO( \pa^4),
\eeqa
we can also show (\ref{oo}).
Note that if we take the other way of ordering,
(\ref{oo}) is not necessarily satisfied.

From (\ref{comp}) and (\ref{oo}),
we conclude that
the noncommutative Dirac-Born-Infeld Lagrangian with 
one of these orderings, $\hat{\cL}_{\p}$, satisfies
the desired equation
\beq
\D \hat{\cL}_{\p}=  {\rm total} \,\, {\rm derivative}
+ \cO( \pa^4).
\eeq
Therefore this Lagrangian without two derivative corrections
is allowed by the equivalence in the approximation of 
neglecting $\cO(\pa^4)$,
but keeping an arbitrary order of $\hF$.
This result is consistent with the calculations
of the effective action for the superstring case \cite{AnTs} 
because in this case there are no two-derivative terms 
in the effective action.

For the bosonic case,
it has been known \cite{Ok}  that 
the known two-derivative corrections
derived from the string four-point
amplitude \cite{AnTs} 
and the $\B$ function in the open string $\sigma$ model \cite{AnTs2}
is not consistent with the equivalence with (\ref{map}).
However 
it can be shown \cite{OkTe} that 
if the mapping of $A$ to $\hA$ (\ref{map}) is modified 
by some field redefinition containing $\T,F$ and $\p$,
the equivalence is consistent with the result 
in \cite{AnTs} and \cite{AnTs2}.

Although this modification should be applied 
for the bosonic case,
we will consider the two-derivative corrections 
which are consistent with the equivalence using (\ref{map})
in the rest of this section.
This is because
these corrections can be added consistently 
in the approximation
even if we modify (\ref{map}) 
and may capture some general structures of the 
effective action of the D-brane.
We will also consider the $2m$-derivative corrections
which are consistent with the equivalence using (\ref{map})
in the approximation of neglecting 
$(2m+2)$-derivative terms in the next section.
For the superstring case,
it is possible that 
(\ref{map})
is valid even if we do not neglect the higher-derivative terms.
Hence it is important that
the determination of these corrections.

Then we will show below that
the two-derivative term
\beq
\hat{\cL}_2= \frac{1}{G_s} \det (G+\hF+\p)^{\half} 
\, L_2,
\label{cl2}
\eeq
satisfies
\beq
\D \hat{\cL}_2=  {\rm total} \,\, {\rm derivative}
+ \cO( \pa^4),
\label{dl}
\eeq
where 
\beqa
L_2 &=& \left\{ a_1 \, {(h_S)}^{mn} {(h_S)}^{iq} {(h_S)}^{jp} 
+a_2 \, {(h_S)}^{mi} {(h_S)}^{nq} {(h_S)}^{jp} \right\}
\hD_m \hF_{ij} \hD_n \hF_{pq}, \CR
{(h_S)}^{ij} &=& \left( \frac{1}{G+\hF+\p} \right)^{ij}_{\rm sym}=
\half \left( \frac{1}{G+\hF+\p} \right)^{ij}+
\half \left( \frac{1}{G-\hF-\p} \right)^{ij} \CR
&=& \left( \frac{1}{G+\hF+\p} G \frac{1}{G-\hF-\p} \right)^{ij},
\eeqa
and $a_1$ and $a_2$ are some constants.
Here it is not required to consider the ordering problem of (\ref{cl2})
because the condition (\ref{dl}) means that
the term is consistent with the equivalence in the
approximation of neglecting $\cO(\pa^4)$.
How these terms are derived is explained in the next section.

If we define a differential operator
\beq
\tilde{\D} = \half \D \T^{kl} \hA_k (\pa_l+\hD_l),
\eeq
the differential of $\hat{\cL}_2$ is written as
\beq
\D \hat{\cL}_2=-\half \D \T^{kl} (\pa_l+\hD_l) 
\left( \hA_k \hat{\cL}_2 \right)
+ \frac{1}{G_s} \det (G+\hF+\p)^{\half} \, (\D+\tilde{\D} ) \,
L_2,
\label{diff2}
\eeq
where the first term of the r.h.s of (\ref{diff2}) is
a total derivative.
Hence we consider the variations of $h_S$ and $\hD \hF$
under $\D+\tilde{\D}$.

From 
\beq
(\D+\tilde{\D} ) \hF_{ij} =-( \hF \D \T \hF)_{ij}+ \cO( \pa^4),
\eeq
it obeys
\beqa
(\D+\tilde{\D} ) {h_S}^{ij} &=& 
-\left( \frac{1}{G+\hF+\p} \left( (G+\p) \D \T (G+\p) 
+(\D+\tilde{\D} ) \hF  \right)
\frac{1}{G+\hF+\p} 
\right)^{ij}_{\rm sym} \CR
&=& -\left( \D \T -\frac{1}{G+\hF+\p} 
\hF \D \T  - \D \T \hF
\frac{1}{G+\hF+\p} 
\right)^{ij}_{\rm sym} + \cO( \pa^4) \CR
&=& \left( h_S (\hF \D \T) + (\D \T \hF) h_S \right)^{ij}
+ \cO( \pa^4).
\label{dh}
\eeqa
Remembering that $[ \D, \pa_m]=0$ and 
that $\hD$ explicitly depends on $\T$ through $*$ product,
we see that 
the commutation relation between $\hD$ and $\D$
is
\beqa
[ \D , \hD_m ] E &=& [ \D, \pa_m]E+i [ E, \D \hA_m]
+\D \T^{kl} \pa_k \hA_m \pa_l E +\cO( \pa^5 E) \CR
&=& i \left[ E , -\half \D \T^{kl} \hA_k (\pa_l \hA_m+\hF_{lm}) 
\right]+\D \T^{kl} \pa_k \hA_m \pa_l E +\cO( \pa^5 E),
\label{a1}
\eeqa
where $E$ is an arbitrary function.
The computation of the commutation relation
between $\hD$ and $\tilde{\D}$ can be carried out straightforwardly
\beqa
[\tilde{\D}  , \hD_m ] E &=& \tilde{\D}
\left( \pa_m E+i [ E, \hA_m] \right)
-\hD_m \left( \half \hA_k \D \T^{kl} (\pa_l +\hD_l) E  \right) \CR
&=& -\half \D \T^{kl} \left( \hD_m \hA_k (\pa_l +\hD_l) E
+ i\hA_k [E, \hF_{ml}-\pa_l \hA_m]\right)+\cO( \pa^5 E).
\label{a2}
\eeqa
After some calculations using (\ref{a1}) and (\ref{a2}), 
we can find a simple result 
\beq
[\D+\tilde{\D}  , \hD_m ] E=-(\hF \D \T)_m^{\;\;\; l} \; (\hD_l E)
+\cO( \pa^5 E),
\label{dde}
\eeq
and then we obtain
\beq
(\D+\tilde{\D} )\hD_m  \hF_{ij} =
-\left( (\hD_m   \hF)  \D \T \hF \right)_{ij}
-\left( \hF \D \T ( \hD_m \hF) \right)_{ij}
-(\hF \D \T)_m^{\;\;\; l} \; (\hD_l \hF_{ij})
+\cO( \pa^5).
\label{ddf}
\eeq
This and (\ref{dh}) imply
that $(\D+\tilde{\D} ) L_2=0$.
Thus we conclude that 
$\D \hat{\cL}_2 ={\rm total} \,\, {\rm derivative} +\cO( \pa^6)$.
However $\cO( \pa^4 )$ terms may exist if the ordinary multiplication 
in (\ref{cl2}) is replaced by $*$ product.
Thus only (\ref{dl}) is meaningful because 
we have not considered the ordering problems of (\ref{cl2}).

Now we discuss the expansion about $F$ of two-derivative corrections
of the effective Lagrangian (\ref{cl2}) 
with $B=\p=\T=0$ and $g_{ij}=\D_{ij}$.
In this commutative description,
the Dirac-Born-Infeld Lagrangian (\ref{Ln}) becomes 
$1/(g_s (2\pi)^{\frac{p+1}{2}}) \sqrt{\det (1+F) }$.
Using $\det(G+\hF+\p)^{\half}=1-\frac{1}{4} \Tr F^2+\cO(F^4)$ and 
$h_S=\frac{1}{1-F^2}=1+ F^2+\cO(F^4)$,
we see 
\beqa
\hat{\cL}_2 &=& a_1 \frac{1}{g_s} \left[
\left(1 -\frac{1}{4} \Tr F^2 \right) \pa_m F_{ij} \pa_m F_{ji}+
(F^2)_{mn} \pa_m F_{ij} \pa_m F_{ji} 
+2 (F^2)_{ik} \pa_m F_{ij} \pa_m F_{jk} \right]  \CR
&&+ a_2\frac{1}{g_s} \left[
\left(1 -\frac{1}{4} \Tr F^2 \right) \pa_m F_{im}  \pa_{n} F_{ni} 
 +2 (F^2)_{mj}  \pa_m F_{ij} \pa_{n} F_{ni}
+(F^2)_{iq} \pa_m F_{im} \pa_{n} F_{nq}
\right] \CR
&&  \hspace{1.5cm} +\cO ( F^4 \pa F \pa F).
\eeqa
Here following \cite{AnTs},
we define a basis of terms of order $F^2 \pa F \pa F$ as
\beqa
&& J_1=F_{kl}F_{lk} \pa_n F_{ij} \pa_n F_{ji},
\hspace{1cm} J_2=F_{kl} F_{li} \pa_n F_{ij} \pa_n F_{jk}, \CR
&& J_3=F_{ni}F_{im} \pa_n F_{kl} \pa_m F_{lk},
\hspace{1cm} J_4=F_{kl} F_{lk} \pa_n F_{ni} \pa_m F_{im}, \CR
&& J_5=-F_{jk}F_{km} \pa_n F_{ni} \pa_m F_{ij},
\hspace{1cm} J_6=F_{kl} F_{lk} \pa_m \pa_m F_{ij} F_{ji}, \CR
&& J_7=\pa_m \pa_m F_{ij} F_{jk}F_{kl}F_{li} .
\eeqa
Therefore after some computations, 
we obtain
\beq
\hat{\cL}_2 = a_1 \frac{1}{g_s} \! \left(\pa_m F_{ij} \pa_m F_{ji}
-\frac{1}{4} J_1 +2 J_2 +J_3 \right)+
a_2 \frac{1}{g_s} \! \left(\half \pa_m F_{ij} \pa_m F_{ji}
-J_5+\frac{1}{8} J_6 -\half J_7 \right)+\cdots,
\label{rco}
\eeq
where the ellipsis represent 
total derivative terms and $\cO(F^4 \pa F \pa F)$ terms.
This is same as the effective Lagrangian obtained in \cite{Ok}.
Thus the result obtained in this paper implies
that the one obtained in \cite{Ok}
is consistent even in the arbitrary order of $F$.
It is noted that in \cite{Ok} only the equivalence without assuming
the existence of $\p$ is required.

\section{General forms of the derivative corrections}

\renewcommand{\theequation}{4.\arabic{equation}}\setcounter{equation}{0}

In this section,
we discuss the other two-derivative corrections
which satisfy (\ref{dl})
and also the higher derivative corrections.

Requiring the noncommutative gauge invariance 
and the gauge invariance for the $B$ field, the most general 
two derivative terms
are 
\beq
\frac{1}{G_s} \det (G+\hF+\p)^{\half} 
\left[ T^{ijklmn}(G^{-1},\hF+\p) \, 
\hD_m \hF_{ij} \hD_n \hF_{kl} 
+T^{ijmn}(G^{-1},\hF+\p)  \,  \hD_m \hD_n \hF_{ij}
\right],
\eeq
where $T^{ijklmn}$ and $T^{ijmn}$ are arbitrary tensors
constructed from $(G^{-1})^{pq}$, $M_{pq} \equiv (\hF+\p)_{pq}$,
$\det G$ and $\det M$.
The tensors $T$ should not depend on $(M^{-1})^{pq}$ because 
it has a singularity at $M=0$.

We can easily generalize the 
invariance problem under $\D$ of two-derivative terms
to the one of higher derivative terms.
Thus below we will look for the Lagrangian $\hat{\cL}_{m}$ 
with $m$-derivative which satisfies the invariant condition
\beq
\D \hat{\cL}_{m}=  {\rm total} \,\, {\rm derivative}
+ \cO( \pa^{m+2}),
\label{dl2}
\eeq
which is the condition consistent with the equivalence
in the approximation of neglecting $\cO(\pa^{m+2})$.
To do this, let us consider 
a $(2n,0)$ tensor $T^{p_1 p_2 \cdots p_{2n}}$ 
constructed from $G^{-1}$ and $M$.
It can be written as
\beq
T^{p_1 p_2 \cdots p_{2n}} =
\sum_{k_1=0}^{\infty} \cdots \sum_{k_n=0}^{\infty} 
C_{\{k\}} \, 
(M^{k_1})^{p_1 p_2} (M^{k_2})^{p_3 p_4} \cdots (M^{k_n})^{p_{2n-1} p_{2n}},
\label{ten}
\eeq
where $C_{\{k\}}$ is some function of the scalars
constructed from $M$ and $G$ and
$(M^{k_i})^{p_{2i-1} p_{2i}}$ means 
$( (G^{-1} M)^{k_i} G^{-1})^{p_{2i-1} p_{2i}}$.
We also consider a $(0,2n)$ tensor 
$\hat{J}_{p_1 p_2 \cdots p_{2n}}$
such that
\beq
\hat{J}_{p_1 p_2 \cdots p_{2n}}=
\left\{ (\hD \cdots \hD \hF) \cdots (\hD \cdots \hD \hF)
 \right\}_{p_1 p_2 \cdots p_{2n}}.
\eeq
For given $n$ and $m$, 
where $m$ is the number of derivative $\hD$ in $\hat{J}$,
there are finite number $s$ of independent 
$\hat{J}_{p_1 p_2 \cdots p_{2n}}$ under the identification
using Bianchi identity.
We note that the total divergence terms should not be used 
for the identification.
Taking a basis of these $\hat{J}^{(i)}$, where $1 \leq i \leq s$,
we will study
the invariance of 
\beq
\hat{\cL}_m=\frac{1}{G_s} \det (G+\hF+\p)^{\half} 
\sum_{i=1}^s \sum_{p_1,\cdots,p_{2n}} 
(T_{(i)})^{p_1 p_2 \cdots p_{2n}}
\, (\hat{J}^{(i)})_{p_1 p_2 \cdots p_{2n}},
\label{dclm}
\eeq
where $(T_{(i)})$ is the tensor of the form
(\ref{ten}) with the coefficients $C^{(i)}_{\{k\}}$.
As like the derivation of (\ref{diff2}), we can show that
\beq
\D \hat{\cL}_m={\rm total} \,\, {\rm derivative}
+ \frac{1}{G_s} \det (G+\hF+\p)^{\half} \, (\D+\tilde{\D} ) \,
\left( \sum_{i=1}^s \sum_{p_1,\cdots,p_{2n}} 
(T_{(i)})^{p_1 \cdots p_{2n}}
\, (\hat{J}^{(i)})_{p_1  \cdots p_{2n}} \right).
\label{dlm}
\eeq
However there is a possibility of 
cancellation between the variations under $\D$
of the terms with different $n$'s.
Note that this cancellation can occur only between
the variations of the terms with $n$ and $n+1$.
We will consider this later.

In order to proceed further,
we require the invariance of $\hat{\cL}_m$ with $\hF=0$ first.
From (\ref{dlm}), we have conditions for the invariance as
\beq
\sum_{i=1}^s \sum_{p_1,\cdots,p_{2n}} 
(\D T_{(i)})^{p_1 p_2 \cdots p_{2n}}
\, (\hat{J}^{(i)})_{p_1  \cdots p_{2n}}=0.
\label{dlm1}
\eeq
It can be shown that
$\D G^{-1}|_{G=1}=- \left( \p \D \T +\D \T \p \right) $
and 
\beqa
\left. \D \left( (G^{-1} \p)^k G^{-1}\right)  \right|_{G=1} 
&=&  - \left( 
\delta \theta \Phi^{k+1}+\Phi \delta \theta \Phi^{k} +\cdots+\Phi^{k+1} \delta \theta 
\right) \CR
&& \,\, + \left( 
\delta \theta \Phi^{k-1}+\Phi \delta \theta \Phi^{k-2} +\cdots+\Phi^{k-1} \delta \theta 
\right), 
\eeqa
where $k \geq 1$.
Here we have set $G=1$ after operating $\D$ 
for notational simplicity.
Therefore to satisfy (\ref{dlm1}), we have to take
\beq
(T_{(i)})^{p_1 \cdots p_{2n}}
=C^{(i)} \, 
(h_S)^{p_1 p_2} (h_S)^{p_3 p_4} \cdots (h_S)^{p_{2n-1} p_{2n}},
\eeq
where $C^{(i)}$ is some function of the scalars
constructed from $M$ and $G$.
We also see that this $C^{(i)}$ should be some constant
since $\D \Tr ( (G^{-1} \p)^{2k} )  |_{G=1} 
=2 k \Tr (\D \T (\p^{2k-1}-\p^{2k+1}))$.

Now we require the condition (\ref{dlm1})
without taking $\hat{F}=0$.
Using (\ref{dh}) and (\ref{ddf}) as in the previous section,
one can easily show that
the condition (\ref{dlm1})
is satisfied for 
\beq
\hat{\cL}_m=\frac{1}{G_s} \det (G+\hF+\p)^{\half} 
\sum_{i=1}^{s_1} C^{(i)} \!\! \sum_{p_1,\cdots,p_{2n}} 
(h_S)^{p_1 p_2} (h_S)^{p_3 p_4} \cdots (h_S)^{p_{2n-1} p_{2n}}
\,\, \bar{J}^{(i)}_{p_1 p_2 \cdots p_{2n}},
\label{clmf}
\eeq
where $\{ \bar{J}^{(i)}_{p_1 p_2 \cdots p_{2n}} \}$ is a basis
of the form 
\beq
((\hD \hF) \cdots (\hD \hF) )_{p_1 p_2 \cdots p_{2n}}.
\eeq
Hence this term is allowed by the equivalence
in the approximation.
In particular, for two-derivative terms,
(\ref{clmf}) is equivalent to (\ref{cl2}).

The terms containing $(\hD)^{N} \hF$ with $N \geq 2$
are other candidates, but 
they can not satisfy the condition (\ref{dlm1}) generically 
because 
there are contributions from  $(\hD)^{N} \hF$ 
which can not be canceled by
the ones from $h_S$ for the variation of $\hat{\cL}_m$ under $\D$
as seen from (\ref{dde}) with $E=(\hD)^{N-1} \hF$.
However, in some cases, 
these are absent because of the symmetry for the indices.
For example, we consider 
\beq
\hat{\cL_2} 
\sim (h_S)^{ip} (h_S)^{jq} [\hD_p, \hD_q] \hF_{ij}.
\label{l2mdd}
\eeq
Remember that
\beq
(\D+\tilde{\D}) (\hD_p \hD_q \hF_{ij})=
-(\D \T)^{lk} \left( (\hD_p\hF_{ql})( \hD_k \hF_{ij})
+(\hD_q\hF_{il} )(\hD_p \hF_{kj})
+(\hD_p\hF_{il} )( \hD_q \hF_{kj} ) \right)
+\cdots,
\label{d+d}
\eeq
where the ellipsis represents the terms which are canceled by the 
contribution from $h_S$ and $\cO(\pa^6)$.
Thus
we obtain $(\D+\tilde{\D}) ((h_S)^{ip} (h_S)^{jq} [\hD_p, \hD_q] \hF_{ij})
=\cO(\pa^6)$ from the Bianchi identity and 
the symmetries of the indices.
Therefore (\ref{l2mdd}) is 
also allowed by the equivalence in the approximation
though this vanishes at $\T=0$.
Note that 
(\ref{l2mdd}) is the only allowed term 
with two-derivative and one $\hF$
because of the Bianchi identity and 
the symmetry for the indices of $\hF_{ij}$.

There are invariant combinations of
the terms with different numbers of indices 
which implies different numbers of $\hF$.
To illustrate these,
we consider 
\beq
\hat{\cL}_4^A 
= \frac{C}{G_s} \det (G+\hF+\p)^{\half} 
(h_S)^{pq} (h_S)^{tu} (h_S)^{ic} (h_S)^{jd} 
(\hD_p \hD_q \hF_{ij} )
(\hD_t \hD_u \hF_{cd} ),
\label{ex}
\eeq
where $C$ is some constant.
In this case, using (\ref{d+d})
the unnecessary terms are easily computed as
\beq
\D \hat{\cL}_4^A \sim -2 \frac{C}{G_s} \det (G+\hF+\p)^{\half}
(h_S)^{pq} (h_S)^{tu} (h_S)^{ic} (h_S)^{jd}
(\D \T)^{lk}  (\hD_p\hF_{ql})( \hD_k \hF_{ij})
(\hD_t \hD_u \hF_{cd} ),
\eeq
where we neglect the terms which are canceled by the 
contribution from $h_S$ and $\cO(\pa^8)$.
Let us define 
\beqa
{(h_A)}^{ij} &=& 
\left( \frac{1}{G+\hF+\p} \right)^{ij}_{\rm anti. sym}=
\left( \frac{1}{G+\hF+\p} (\hF+\p)  \frac{1}{G-\hF-\p} \right)^{ij},
\eeqa
which obeys
\beqa
(\D+\tilde{\D} ) {h_A}^{ij} &=& 
-\left( \frac{1}{G+\hF+\p} \left( (G+\p) \D \T (G+\p) 
+(\D+\tilde{\D} ) \hF  \right)
\frac{1}{G+\hF+\p} 
\right)^{ij}_{\rm anti. sym} \CR
&=& -(\D \T)^{ij} +\left( h_A (\hF \D \T) + (\D \T \hF) h_A \right)^{ij}
+ \cO( \pa^4).
\label{dha}
\eeqa
Then it can be seen that
\beq
\D (\hat{\cL}_4^A+\hat{\cL}_4^B+\hat{\cL}_4^C) \sim 0,
\eeq
where
\beqa
\hat{\cL}_4^B
&=& -2 \frac{C}{G_s} \det (G+\hF+\p)^{\half} 
(h_S)^{pq} (h_S)^{tu} (h_S)^{ic} (h_S)^{jd}
(h_A)^{lk}  (\hD_p\hF_{ql})( \hD_k \hF_{ij})
(\hD_t \hD_u \hF_{cd} ), \CR
\hat{\cL}_4^C
&=& \;\;\;\; \frac{C}{G_s} \det (G+\hF+\p)^{\half} 
(h_S)^{pq} (h_S)^{tu} (h_S)^{ic} (h_S)^{jd}
(h_A)^{lk} (h_A)^{ef}  \CR
&& \hspace{6cm} \times (\hD_p\hF_{ql})( \hD_k \hF_{ij})
(\hD_t\hF_{ue}) (\hD_f \hF_{cd} ). 
\eeqa
As this example,
for general terms of the form 
\beq
\hat{\cL}_m=\frac{1}{G_s} \det (G+\hF+\p)^{\half} 
\sum_{i=1}^{s_1} C^{(i)} \!\! \sum_{p_1,\cdots,p_{2n}} 
(h_S)^{p_1 p_2} (h_S)^{p_3 p_4} \cdots (h_S)^{p_{2n-1} p_{2n}}
\,\, \hat{J}^{(i)}_{p_1 p_2 \cdots p_{2n}},
\label{clmf2}
\eeq
we may construct the invariant combinations
by adding certain terms like $\hat{\cL}^B_4$ and $\hat{\cL}^C_4$.

Therefore we conclude that the general forms of 
the allowed $m$-derivative corrections in the approximation
of neglecting $\cO(\pa^{m+2})$
are given by (\ref{clmf}) and (\ref{clmf2}) with 
certain terms like $\hat{\cL}^B_4$ and $\hat{\cL}^C_4$.

Finally we study the behavior of the derivative corrections at $\T=0$
in the zero slope limit of \cite{SeWi}, $\A' \sim \E^{\half}$,
$g_{ij}=\E \D_{ij}$ with $\E \rightarrow 0$.
In \cite{SeWi} it has been shown that
\beq
\cL_{DBI}=\frac{1}{g_s (2 \pi)^p (\A')^{\frac{p+1}{2}}}
\left( | {\rm Pf(F+B)} | 
-\frac{\E^2}{4} | {\rm Pf(F+B)} | \Tr \frac{1}{(F+B)^2}
+\cO(\E^3) \right),
\eeq
where the first term is a constant plus a total derivative.
We can show that 
\beq
h_S = -\frac{\E}{ (2 \pi \A')^2} \frac{1}{(F+B)^2} \sim \cO(\E^0),
\eeq
and 
\beq
h_A =
-\frac{1}{ 2 \pi \A'} \frac{1}{(F+B)} \sim \cO(\E^{-\frac{1}{2}}).
\eeq
From the dimensional analysis,
the constants $C^{(i)}$ and $C$ in (\ref{clmf}) and (\ref{clmf2}),
respectively, are restricted.
Indeed, we see that $C$ or $C^{(i)} \sim \A'^{-(p+1)/2+ n_S+ n_A}$,
where $n_S$ and $n_A$ are the number of the $h_S$ and $h_A$
in the Lagrangians $\hat{\cL}$, respectively.
Thus
\beqa
\hat{\cL} & \sim & 
\frac{1}{g_s (2 \pi)^p (\A')^{\frac{p+1}{2}}} \CR
 & & \times 
\left( \E^{\frac{n_S}{2}} | {\rm Pf(F+B)} | 
\frac{1}{(F+B)^{2 n_S}} 
\frac{1}{(F+B)^{n_A} }
J_{p_1 p_2 \cdots p_{2 (n_S+n_A)}}
+\cO(\E^{\frac{n_S}{2}+2}) \right),
\eeqa
where $J$ is $\hat{J}^{(i)}$ or $\tilde{J}$ with $\hD=\pa$ and $\hF=F$.
This is negligible compared with $\cL_{DBI}$ if $n_s > 4$.
Therefore for the superstring case,
the only remaining derivative corrections 
in the limit are
the terms like $\hat{\cL}_4^A+\hat{\cL}_4^B+\hat{\cL}_4^C$.
This result may have application for a deeper understanding 
of the relation between the instanton on the noncommutative space
\cite{NeSh}
and the instanton solution
in the Dirac-Born-Infeld Lagrangian 
with nonzero $B$ field \cite{Te,SeWi, MaMiMoSt}.

\section{Conclusion}

\renewcommand{\theequation}{5.\arabic{equation}}\setcounter{equation}{0}

We have considered
the derivative corrections to the Dirac-Born-Infeld action
consistent with the equivalence  
between the noncommutative gauge theories and
the ordinary gauge theory.
In particular, we have shown that in the approximation of 
neglecting the fourth and higher order derivative terms
the D-brane action computed in the superstring theory
is consistent with the equivalence.

We have also explicitly constructed
the general forms of 
the $2n$-derivative corrections
which satisfy this equivalence relation
in the approximation of neglecting the $(2n+2)$-derivative terms.
It may capture some general structures of the 
effective action of the D-brane.

It is interesting to generalize
the results obtained in this paper to
the effective theories on several D-branes.
In this case, we should treat the non-Abelian gauge fields,
so that the ordering problems exist even for 
the ordinary gauge fields which have not been solved yet.
Thus the constraints using the equivalence 
are expected to be important for 
determination of the effective action on the several D-branes.

\vskip6mm\noindent
{\bf Acknowledgements}

\vskip2mm
I would like to thank J. Hashiba, K. Hosomichi
and K. Okuyama for useful discussions.
I would also like to thank K. Hashimoto, T. Kawano and Y. Okawa
for discussions on the derivative corrections.
This work was supported in part by JSPS Research Fellowships for Young 
Scientists. \\

\noindent
{\bf Note added}: 

As this article was being completed,
we received the preprint
\cite{Co} which give 
the derivative corrections for 
the Dirac-Born-Infeld Lagrangian
which is invariant under a simplified version \cite{CoSc,Is}
of the Seiberg-Witten map.
The two-derivative correction obtained in \cite{Co} 
\begin{equation}
g_{kl} g_{pq} \left( \frac{1}{g+F} \right)^{ij} 
\partial_i  \left( \frac{1}{g+F} \right)^{kp}  
\partial_i \left( \frac{1}{g+F} \right)^{lq}   
= -(h_S)^{ij} {\rm Tr} \left( h_S (\partial_i F) h_S (\partial_j F) \right),
\end{equation} 
coincides with the $\hat{L}_2 $ obtained in this paper with $a_1=-1$ and 
$a_2=0$.
Note that on the computation of the expansion about $F$,
some terms are omitted in the eq.(4) in \cite{Co}.


\newpage


\end{document}